\documentclass{elsart}
\usepackage{natbib}
\usepackage{amsfonts,amssymb,amsmath,amsbsy}
\usepackage{epsfig,graphicx}

\newtheorem{theo}{Theorem}[section]

\newtheorem{lema}[theo]{Lemma}
\newtheorem{example}[theo]{Example}
\newtheorem{defi}[theo]{Definition}

\newtheorem{clai}[theo]{Claim}
\newtheorem{conjec}[theo]{Conjecture}
\newenvironment{proof}{\noindent{\bf Proof:}}{\hfill $\Box$}

\def\build#1_#2^#3{\mathrel{\mathop{\kern 0 pt#1}\limits_{#2}^{#3}}}

\newcommand{\ud}{\,\mathrm{d}}
\newcommand{\cN}{\mathcal{N}}
\newcommand{\cD}{\mathcal{D}}
\newcommand{\cA}{\mathcal{A}}
\newcommand{\cB}{\mathcal{B}}

\newcommand{\ba}{\mathbf{a}}
\newcommand{\be}{ \mbox{\boldmath$\epsilon$}}
\begin{document}

\begin{frontmatter}
\title{Pairs of SAT Assignment\\ in Random Boolean Formul\ae}
\journal{Theoretical Computer Science}

\author{Herv\'e Daud\'e}
\address{LATP, UMR 6632 CNRS et Universit\'e de Provence\\
13453 Marseille CEDEX, France} 
\author{Marc M\'ezard}
\address{LPTMS, UMR 8626 CNRS et Universit\'e Paris-Sud\\
 91405 Orsay CEDEX, France}
\author{Thierry Mora}
\address{LPTMS, UMR 8626 CNRS et Universit\'e Paris-Sud\\
 91405 Orsay CEDEX, France}
\author{Riccardo Zecchina}
\address{Physics Department, Politecnico di Torino\\
Corso Duca degli Abruzzi 24, 10129 Torino, Italy}

%\date{\today}

\begin{abstract}
We investigate geometrical properties of the random $K$-satisfiability
problem using the notion of $x$-satisfiability: a formula is $x$-satisfiable is there exist two SAT assignments differing in $Nx$ variables. We show the existence of a sharp threshold for this property as a function of the clause density. For large enough $K$, we prove that there exists a region of clause
density, below the satisfiability threshold, where the landscape of Hamming distances between SAT assignments experiences a gap: pairs of SAT-assignments exist at small $x$, and around $x=\frac{1}{2}$, but they do not exist at intermediate values of $x$.
This result is consistent with the clustering
scenario which is at the heart of the recent heuristic
analysis of satisfiability using statistical physics analysis (the cavity
method), and its algorithmic counterpart (the survey propagation algorithm).
Our method uses elementary probabilistic arguments (first and second moment
methods), and might be useful in other problems of computational and
physical interest where similar phenomena appear.

\end{abstract}
\end{frontmatter}

%\maketitle

\section{Introduction and outline}

Consider a string of Boolean variables ---\,or equivalently a string of
\emph{spins}\,--- of size $N$: $\vec\sigma=\{\sigma_i\}\in\{-1,1\}^{N}$. Call
a $K$-clause a disjunction binding $K$ of these Boolean variables in such a
way that one of their $2^K$ joint assignments is set to {\sc false}, and all
the others to {\sc true}. A formula in a conjunctive normal form (CNF) is a
conjunction of such clauses. The  satisfiability problem is 
stated as: does there exist a truth assignment $\vec\sigma$ that satisfies this
formula? A  CNF formula is said to be \emph{satisfiable} (SAT) if this is the
case, and \emph{unsatisfiable} (UNSAT) otherwise.

The satisfiability problem is often viewed as the canonical constraint
satisfaction problem (CSP). It is  the first problem to have been
shown NP-complete \cite{cook}, i.e. at least as hard as any problem for which
a solution can be checked in polynomial time.

The $P\neq NP$ conjecture states that no general polynomial-time algorithm exists that can
decide whether a formula is SAT or UNSAT. However formulas which are encountered in practice
can often be solved easily. In order to understand properties of some typical families
of formulas, one introduces a probability measure on the set of instances.
In  the random
$K$-SAT problem, one generates   a random $K$-CNF formula $F_K(N,M)$ as a
conjunction of $M=N\alpha$ $K$-clauses, each of them being uniformly drawn
from the $2^K\binom{N}{K}$ possibilities. In the recent years the random
$K$-satisfiability problem has attracted much interest in computer science and
in statistical physics. Its most striking feature is certainly its sharp
threshold.

Throughout this paper, `with high probability' (w.h.p.) means with a probability which goes to one as $N \to \infty$.
\begin{conjec}[Satisfiability Threshold Conjecture]\label{stc}
For all $K\geq 2$, there exists $\alpha_c(K)$ such that:
\begin{itemize}
\item if $\alpha<\alpha_c(K)$, $F_K(N,N\alpha)$ is satisfiable w.h.p.
\item if $\alpha>\alpha_c(K)$, $F_K(N,N\alpha)$ is unsatisfiable w.h.p.
\end{itemize}
\end{conjec}
The random $K$-SAT problem, for $N$ large and $\alpha$ close to $\alpha_c(K)$, provides instances of very hard CNF formulas that can be used as benchmarks for algorithms. For such hard ensembles, the study of the typical complexity could be crucial for the understanding of the usual `worst-case' complexity.

Although Conjecture \ref{stc} remains unproved, Friedgut established the existence of a non-uniform sharp threshold \cite{Friedgut}.
\begin{theo}[Friedgut]\label{fried1}
For each $K\geq 2$, there exists a sequence $\alpha_N(K)$ such that for all $\epsilon>0$:
\begin{equation}
\lim_{N\to\infty}\mathbf{P}(F_K(N,N\alpha)\textrm{ is satisfiable})=
\left \{  \begin{array}{ll}1 & \textrm{if } \alpha=(1-\epsilon)\alpha_N(K) \\ 0 & \textrm{if } \alpha=(1+\epsilon)\alpha_N(K).\end{array} \right.
\end{equation}
\end{theo}

A lot of efforts have been devoted to finding tight bounds for the threshold.
The best upper bounds so far were derived using first moment methods
\cite{kirousis,dubois}, and the best lower bounds were obtained by second
moment methods \cite{achliomoore,achlioperes}. Using these bounds, it was
shown that $\alpha_c(K)=2^K \ln(2)-O(K)$ as $K\to\infty$.

On the other hand, powerful, self-consistent, but non-rigorous tools from 
statistical physics were used to predict specific values of $\alpha_c(K)$, as
well as heuristical asymptotic expansions for large $K$ \cite{MZ,MPZ,MMZ-RSA}.
The \emph{cavity method} \cite{Cavity}, which provides these results,
relies on several unproven assumptions motivated by spin-glass theory, the most
important of which is the partition of the space of SAT-assignments into many
\emph{states} or \emph{clusters} far away from each other (with Hamming distance greater than $cN$ as $N\to\infty$), in the so-called hard-SAT phase. 

So far, the existence of such a clustering
phase has been shown rigorously in the simpler case of the random XORSAT
problem \cite{XORSAT-CDMM,XORSAT-MRZ,XORSAT-DM} in
compliance with the prediction of the cavity method, but its existence is predicted in many other problems, such as $q$-colorability \cite{mulet,braunstein} or the Multi-Index Matching Problem \cite{martinmezardrivoire}. 
At the heuristic level, clustering is an important phenomenon,
often held
responsible for entrapping local search algorithm into non-optimal metastable
states \cite{montanarisemerjian}. It is also a limiting feature for the belief propagation iterative decoding
algorithms in Low Density Parity Check Codes
\cite{montanari,FLMR}. 

In this paper we provide a rigorous analysis of some geometrical properties of the space of SAT-assignments in the random $K$-SAT problem. This study complements the results of \cite{MMZ_prl}, and its results are consistent with the clustering
scenario. A new characterizing feature of CNF formulas, the
`$x$-satisfiability', is proposed, which carries information about the
spectrum of distances between SAT-assignments. The $x$-satisfiability property is studied thoroughly using first and second moment methods previously developed for the satisfiability
threshold.

\bigskip
The Hamming distance between two assignments  $(\vec \sigma,\vec \tau)$ is defined by
\begin{equation}
d_{\vec\sigma\vec\tau}=\frac{N}{2}-\frac{1}{2}\sum_{i=1}^{N}\sigma_i\tau_i \ .
\end{equation}
(Throughout the paper the term `distance' will always refer to the Hamming distance.)
Given a random formula $F_K(N,N\alpha)$, we define a `SAT-$x$-pair' as a 
pair of assignments $(\vec \sigma,\vec \tau)\in\{-1,1\}^{2N}$, which both satisfy $F$, and which are 
at a fixed distance specified by $x$ as follows:
\begin{equation}\label{eq:condepsilon}
d_{\vec\sigma\vec\tau} \in [Nx-\epsilon(N),Nx+\epsilon(N)].
\end{equation}
Here $x$ is the proportion of distinct values between the two configurations, which we keep fixed as $N$ and $d$ go to infinity. The resolution $\epsilon(N)$ has to be $\geq 1$ and sub-extensive: $\lim_{N\rightarrow\infty}\epsilon(N)/N=0$, but its precise form is unimportant for our large $N$ analysis. For example we can choose $\epsilon(N)=\sqrt{N}$.
\begin{defi}
A CNF formula is $x$-satisfiable if it possesses a SAT-$x$-pair.
\end{defi}
Note that for $x=0$, $x$-satisfiability is equivalent to satisfiability, while for $x=1$, it is equivalent to Not-All-Equal satisfiability, where each clause must contain at least one satisfied literal and at least one unsatisfied litteral \cite{gareyjohnson}.

The clustering property found heuristically in  \cite{MPZ,MZ} suggests the following:
\begin{conjec}\label{cluster}
For all $K\geq K_0$, there exist $\alpha_1(K)$, $\alpha_2(K)$, with $\alpha_1(K)<\alpha_2(K)$, such that:
 for all $\alpha\in(\alpha_1(K),\alpha_2(K))$, there exist $x_1(K,\alpha)<x_2(K,\alpha)<x_3(K,\alpha)$ such that:
\begin{itemize}
\item
for all $x\in [0,x_1(K,\alpha)]\cup[x_2(K,\alpha),x_3(K,\alpha)]$, a random formula $F_K(N,N\alpha)$ is $x$-satisfiable w.h.p.
\item for all  $x \in [x_1(K,\alpha),x_2(K,\alpha)]\cup [x_3(K,\alpha),1]$, a random formula $F_K(N,N\alpha)$ is $x$-unsatisfiable w.h.p.
\end{itemize}
\end{conjec}
Let us give a geometrical interpretation of this conjecture. The space of SAT-assignments is partioned into non-empty regions whose diameter is smaller than $x_1$; the distance between any two of these regions is at least $x_2$, while $x_3$ is the maximum distance between any pair of SAT-assignments. 
This interpretation is compatible with the notion of clusters used in the statistical physics approach.
It should also be mentioned that in a contribution posterior to this work \cite{AchlioptasRicci06}, the number of regions was shown to be exponential in the size of the problem, further supporting the statistical mechanics picture.

Conjecture \ref{cluster} can be rephrased  in a slightly different way, 
which decomposes it into two steps. The first step is to state the
 \emph{Satisfiability Threshold Conjecture} for pairs:
\begin{conjec}\label{sharpxsat}
For all $K\geq 2$ and for all $x$, $0<x<1$, there exists an $\alpha_c(K,x)$ such that:
\begin{itemize}
\item if $\alpha<\alpha_c(x)$, $F_K(N,N\alpha)$ is $x$-satisfiable w.h.p.
\item if $\alpha>\alpha_c(x)$, $F_K(N,N\alpha)$ is $x$-unsatisfiable w.h.p.
\end{itemize}
\label{x-stconj}
\end{conjec}
The second step conjectures that for $K$ large enough, as a function of $x$, the function  $\alpha_c(K,x)$ is
non monotonic and has two maxima: a local maximum at a value $x_M(K)<1$, and a global maximum at $x=0$.

\bigskip
In this paper we prove the equivalent of  Friedgut's theorem:
\begin{theo}\label{fried2}
For each $K\geq 3$ and $x$, $0<x<1$, there exists a sequence $\alpha_N(K,x)$ such that for all $\epsilon>0$:
\begin{equation}
\lim_{N\to\infty}\mathbf{P}(F_K(N,N\alpha)\textrm{ is }x\textrm{-satisfiable})=
\left \{  \begin{array}{ll}1 & \textrm{if } \alpha=(1-\epsilon)\alpha_N(K,x), \\ 0 & \textrm{if } \alpha=(1+\epsilon)\alpha_N(K,x),\end{array} \right.
\end{equation}
\end{theo}

and we obtain two functions, $\alpha_{LB}(K,x)$ and $\alpha_{UB}(K,x)$, such that:

\begin{itemize}
\item For $\alpha>\alpha_{UB}(K,x)$, a random $K$-CNF $F_K(N,N\alpha)$ is  $x$-unsatisfiable w.h.p.
\item  For $\alpha<\alpha_{LB}(K,x)$, a random $K$-CNF $F_K(N,N\alpha)$ is  $x$-satisfiable w.h.p.
\end{itemize}

\begin{figure}
\begin{center}
\resizebox{.9\linewidth}{!}{\includegraphics{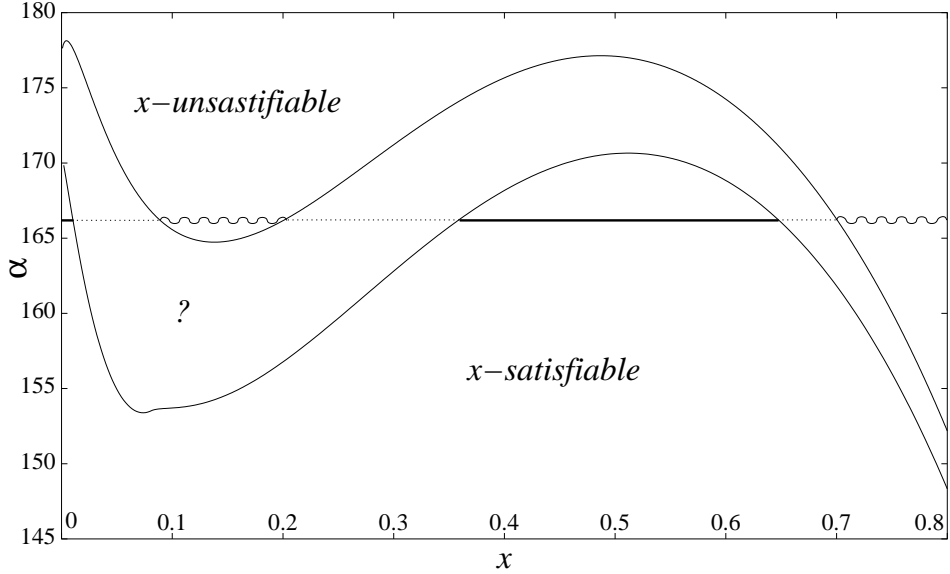}}
\caption{\label{alpha8}\small Lower and Upper Bounds for $\alpha_N(K=8,x)$. 
The Upper Bound is obtained by the first moment method. Above this curve
 there exists no SAT-$x$-pair, w.h.p. The Lower Bound is obtained by the second moment method.
Below this curve there exist a SAT-$x$-pair w.h.p. For $164.735<\alpha<170.657$, these curves confirm the existence of a clustering phase, illustrated here for $\alpha=166.1$: solid lines represent $x$-sat regions, and wavy lines $x$-unsat regions. The $x$-sat zone near $0$ corresponds to SAT-assignments belonging to the same region, whereas the $x$-sat zone around $\frac{1}{2}$ corresponds to SAT-assignments belonging to different regions. The $x$-unsat region around $.13$ corresponds to the inter-cluster gap. We recall that the
best refined lower and upper bounds for the satisfiability threshold
$\alpha_c(K=8)$ from \cite{dubois,achlioperes} are respectively $173.253$ and
$176.596$. The cavity prediction is $176.543$ \cite{MMZ-RSA}.
}
\end{center}
\end{figure}

The two functions
  $\alpha_{LB}(K,x)$ and $\alpha_{UB}(K,x)$ are lower and upper bounds for $\alpha_N(K,x)$ as $N$ tends to infinity. 
Numerical computations of these bounds indicate  that $\alpha_N(K,x)$ is
non monotonic as a function of $x$ for $K\geq 8$, as illustrated in Fig.~\ref{alpha8}.
More precisely, we prove
\begin{theo}\label{nonmonotonic}
For all $\epsilon>0$, there exists $K_0$ such that for all $K\geq K_0$,
\begin{eqnarray}
\min_{x\in\left(0,\frac{1}{2}\right)} \alpha_{UB}(K,x)& \leq & (1+\epsilon) \frac{2^{K}\ln 2}{2},\label{ineq1}\\
\alpha_{LB}(K,0) & \geq & (1-\epsilon) 2^{K}\ln 2,\label{ineq2}\\
\alpha_{LB}(K,1/2) & \geq & (1-\epsilon) 2^{K}\ln 2.\label{ineq3}
\end{eqnarray}
\end{theo}

This in turn shows that, for $K$ large enough and in some well chosen interval
of $\alpha$ below the satisfiability threshold $\alpha_c\sim 2^K\ln 2$, SAT-$x$-pairs exist
for $x$ close to zero and for $x=\frac{1}{2}$, but they do not exist in the intermediate $x$ region. Note that Eq.~\eqref{ineq2} was established by \cite{achlioperes}.

In section \ref{first} we establish rigorous and explicit upper bounds using
the first-moment method. The existence of a gap interval is proven in a certain range of $\alpha$, and bounds on this interval are found, which imply Eq.~\eqref{ineq1} in Theorem \ref{nonmonotonic}.
Section \ref{second} derives the lower bound, using a
weighted second-moment method, as developed recently in 
\cite{achliomoore,achlioperes}, and presents numerical results. In section \ref{largek} we discuss the behavior of the
lower bound for large $K$. The case of $x=\frac{1}{2}$ is treated rigorously, and Eq.~\eqref{ineq3} in Theorem \ref{nonmonotonic} is proven. Other values of $x$ are treated at the heuristic level. Section
\ref{prooffriedgut} presents a proof of Theorem \ref{fried2}. We discuss our
results in section \ref{conclu}.

\section{Upper bound: the first moment method}\label{first}
The first moment method relies on Markov's inequality:
\begin{lema}\label{lemmafirst}
Let $X$ be a non-negative random variable. Then
\begin{equation}
\mathbf{P}(X\geq 1)\leq \mathbf{E}(X)\ .
\label{lemfirst}
\end{equation}
\end{lema}

We take $X$ to be the number of pairs of SAT-assignments at fixed distance:
\begin{equation}\label{Z1}
Z(x,F)=\sum_{\vec \sigma,\vec \tau}\delta\left(d_{\vec\sigma\vec\tau}\in[Nx+\epsilon(N),Nx-\epsilon(N)]\right)
\;\delta\left[\vec\sigma,\,\vec\tau \in S(F)\right],
\end{equation}
where $F=F_K(N,N\alpha)$ is a random $K$-CNF formula, and $S(F)$ is the set of SAT-assignments to this formula. 
Throughout this paper $\delta(A)$ is an indicator function, equal to $1$ if the statement $A$ is true, equal to $0$
otherwise. The expectation $\mathbf{E} $ is over the set of random $K$-CNF formulas.
Since $Z(x,F)\geq 1$ is equivalent to `$F$ is $x$-satisfiable', (\ref{lemfirst})  gives  an upper 
bound for the probability of $x$-satisfiability.

The expected value of the double sum can be rewritten as:
\begin{equation}
\mathbf{E}(Z)=2^N \sum_{d\in[Nx+\epsilon(N),Nx-\epsilon(N)]\cap \mathbb{N}}\binom{N}{d}\mathbf{E}\left[\delta\left(\vec\sigma,\vec\tau\in S(F)\right)\right].
\end{equation}
where $\vec\sigma$ and $\vec\tau$ are any two assignments with Hamming distance $d$.
We have $\delta\left(\vec\sigma,\vec\tau\in S(F)\right)=\prod_c \delta\left(\vec\sigma,\vec\tau\in S(c)\right)$, where $c$ denotes one of the $M$ clauses. All clauses are drawn independently, so that we have:
\begin{equation}
\mathbf{E}(Z)\leq (2\epsilon(N)+1)2^N \max_{d\in[Nx+\epsilon(N),Nx-\epsilon(N)]\cap \mathbb{N}}\left\{\binom{N}{d}\left({\mathbf{E}\left[\delta\left(\vec\sigma,\vec\tau\in S(c)\right)\right]}\right)^{M}\right\},
\end{equation}
where we have bounded the sum by the maximal term times the number of terms.
$\mathbf{E}\left[\delta\left(\vec\sigma,\vec\tau\in S(c)\right)\right]$ can easily be calculated and its value is: $1-2^{1-K}+2^{-K}(1-x)^{K}+o(1)$. Indeed there are only two realizations of the clause among $2^K$ that do not satisfy $c$ unless the two configurations overlap exactly on the domain of $c$.

Considering the normalized logarithm of this quantity, 
\begin{equation}
F(x,\alpha)=\lim_{N\to\infty}\frac{1}{N}\ln \mathbf{E}(Z)=\ln 2+H_2(x)+\alpha\ln\left(1-2^{1-K}+2^{-K}(1-x)^{K}\right),
\end{equation}
where $H_2(x)=-x\ln x-(1-x)\ln(1-x)$ is the two-state entropy function, one can deduce an upper bound for 
$\alpha_N(K,x)$. Indeed, $F(x,\alpha)<0$ implies $\lim_{N\to\infty}\mathbf{P}(Z(x,F)\geq 1)=0$.
Therefore:
\begin{theo}\label{thUB}
For each $K$ and $0<x<1$, and for all $\alpha$ such that
\begin{equation}\label{alUB}
\alpha>\alpha_{UB}(K,x)=-\frac{\ln 2+H_2(x)}{\ln(1-2^{1-K}+2^{-K}(1-x)^{K})},
\end{equation}
a random formula
$F_K(N,N\alpha)$ is $x$-unsatisfiable w.h.p.
\end{theo}
We observe numerically that a `gap' ($x_1,x_2$ and $\alpha$ such that $x_1<x<x_2\Longrightarrow F(x,\alpha)< 0$) appears for $K\geq 6$. More generally, the following results holds, which implies Eq.~\eqref{ineq1} in Theorem \ref{nonmonotonic}:
\begin{theo}\label{thgap}
Let $\epsilon\in (0,1)$, and $\{y_K\}_{K\in \mathbb{N}}$ be a sequence verifying $Ky_K\to\infty$ and $y_K=o(1)$. Denote by $H_2^{-1}(u)$ the smallest root to $H_2(x)=u$, with $u\in[0,\ln 2]$.

There exists $K_0$ such that for all $K\geq K_0$, $\alpha\in[(1+\epsilon)2^{K-1}\ln 2,\alpha_N(K))$ and $x\in [y_K,H_2^{-1}(\alpha 2^{1-K}-\ln 2-\epsilon)]\cup[1-H_2^{-1}(\alpha 2^{1-K}-\ln 2-\epsilon),1]$, $F_K(N,N\alpha)$ is $x$-unsatisfiable w.h.p.
\end{theo}
{\it Proof. }Clearly $(1+\epsilon)2^{K-1}\ln(2)<\alpha_N(K)$ since $\alpha_N(K)=2^K \ln(2)-O_K(K)$ \cite{achlioperes}. Observe that $(1-y_K)^K =o(1)$. Then for all $\delta>0$, there exists $K_1$ such that for all $K\geq K_1$, $x>y_K$:
\begin{equation}
\alpha_{UB}(x)<(1+\delta)2^{K-1}(\ln 2+H_2(x)).
\end{equation}
Inverting this inequality yields the theorem. \hfill $\Box$

\begin{figure}
\begin{center}
\resizebox{.9\linewidth}{!}{\epsfig{file=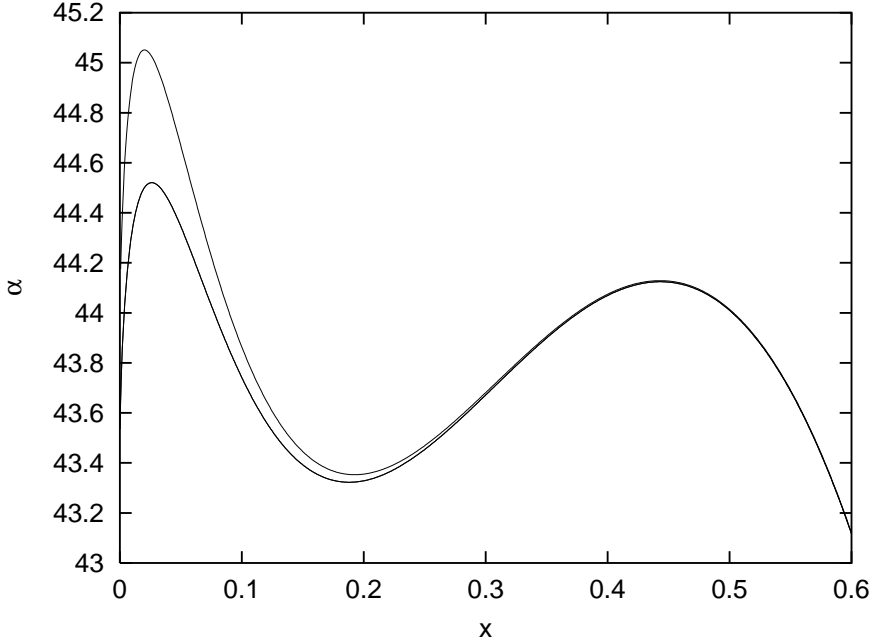,width=.8\linewidth,angle=-90}}
\caption{\label{upper} \small Comparison between the simple upper bound \eqref{alUB} for $\alpha_{N}(K=6,x)$ (top curve) and the refined one (bottom curve), as defined in Theorem \ref{thref1}.}
\end{center}
\end{figure}

The choice (\ref{Z1}) of $X$, although it is the simplest one, is not optimal. The first moment method only requires the condition $X\geq 1$ to be equivalent to the $x$-satisfiability, and better choices of $X$
exist which allow to improve the bound. Techniques similar to the one introduced separately by
Dubois and Boufkhad \cite{dubois} on the one hand, and Kirousis, Kranakis and
Krizanc \cite{kirousis} on the other hand, can be used to obtain two tighter bounds.
Quantitatively, it turns out that these more elaborate bounds provide only very little improvement over the simple bound  \eqref{alUB} (see
Fig.~\ref{upper}). For the sake of completeness, we give without proof the simplest of these bounds:
\begin{theo}\label{thref1}The unique positive solution of the equation
\begin{align}
H_2(x)&+\alpha\ln\left(1-2^{1-K}+2^{-K}(1-x)^{K}\right)\nonumber\\
&+(1-x)\ln\left[2-\exp\left(-K\alpha\frac{2^{1-K}-2^{-K}(1-x)^{K-1}}{1-2^{1-K}+2^{-K}(1-x)^{K}}\right)\right]\nonumber\\
&+x\ln\left[2-\exp\left(-K\alpha\frac{2^{1-K}-2^{1-K}(1-x)^{K-1}}{1-2^{1-K}+2^{-K}(1-x)^{K}}\right)\right]=0
\end{align}
is an upper bound for $\alpha_N(K,x)$. For $x=0$ we recover the expression of \cite{kirousis}.
\end{theo}

The proof closely follows that of \cite{kirousis} and presents no notable difficulty.
We also derived a tighter bound based on the technique used in \cite{dubois}, gaining only a small improvement over the bound of Theorem \ref{thref1} (less than $.001\%$).

\section{Lower bound: the second moment method}\label{second}
The second moment method  uses the following consequence of Chebyshev's inequality:
\begin{lema}\label{lemmasecond}
If $X$ is a non-negative random variable, one has:
\begin{equation}
\mathbf{P}(X>0)\geq \frac{\mathbf{E}(X)^2}{\mathbf{E}(X^2)}.
\end{equation}
\end{lema}
It is well known that the simplest choice of  $X$ as 
 the number of  SAT-assignments (in our case the number of SAT-$x$-pairs) is bound to fail. The intuitive reason \cite{achliomoore,achlioperes} is that this naive choice favors pairs of SAT-assignments with a great number of satisfying litterals. It turns out that such assignments are highly correlated, since they tend to agree with each other, and this causes the failure of the second-moment method. In order to deal with {\it balanced} (with approximately half of literals satisfied) and uncorrelated pairs of assignments, one must consider a weighted sum of all SAT-assignments. Following \cite{achliomoore,achlioperes}, we define:
\begin{equation}
Z(x,F)=\sum_{\vec\sigma,\vec\tau}\delta\left(d_{\vec\sigma\vec\tau}=\lfloor Nx\rfloor\right)W(\vec\sigma,\vec\tau,F),
\end{equation}
where $\lfloor Nx \rfloor$ denotes the integer part of $Nx$. Note that the condition $d_{\vec\sigma\vec\tau}=\lfloor Nx\rfloor$ is stronger than Eq.~\eqref{eq:condepsilon}.
The weights $W(\vec\sigma,\vec\tau,F)$ are decomposed according to each clause:
\begin{eqnarray}
W(\vec\sigma,\vec\tau,F)&=&\prod_c W(\vec\sigma,\vec\tau,c),\\
\textrm{with}\quad W(\vec\sigma,\vec\tau,c)&=&W(\vec u,\vec v),
\end{eqnarray}
where $\vec u,\vec v$ are $K$-component vectors such that: $u_i=1$ if the $i^\textrm{th}$ litteral of $c$ is satisfied under $\vec\sigma$, and $u_i=-1$ otherwise (here we assume that the variables connected to $c$ are arbitrarily ordered). $\vec v$ is defined in the same way with respect to $\vec \tau$.
In order to have the equivalence between $Z>0$ and the existence of pairs of SAT-assignments, we impose the following condition on the weights:
\begin{equation}
W(\vec u,\vec v)=\left\{\begin{array}{ll} 0&\textrm{if}\quad\vec u=(-1,\ldots,-1)\quad\textrm{or}\quad\vec v=(-1,\ldots,-1),\\
>0 &\textrm{otherwise}.\end{array}\right.
\end{equation}

Let us now compute the first and second moments of $Z$:
\begin{clai}
\begin{equation}
\mathbf{E}(Z)=2^N\binom{N}{\lfloor Nx\rfloor}f_1(x)^M,
\end{equation}
where
\begin{eqnarray}
f_1(x)&=&\mathbf{E}[W(\vec\sigma,\vec\tau,c)]\\
&=&2^{-K}\sum_{\vec u,\vec v}W(\vec u,\vec v)(1-x)^{|\overrightarrow{u\cdot v}|}x^{K-|\overrightarrow{u\cdot v}|}.
\end{eqnarray}
Here $|\vec u|$ is the number of indices $i$ such that $u_i=+1$, and $\overrightarrow{u\cdot v}$ denotes the vector $(u_1 v_1,\ldots,u_K v_K)$.
\end{clai}

Writing the second moment is a little more cumbersome:
\begin{clai}
\begin{equation}\label{E2}
\mathbf{E}(Z^2)=2^N\sum_{\mathbf{a}\in V_N\cap \{0,1/N,2/N,\ldots,1\}^8}\frac{N!}{\prod_{i=0}^{7}(Na_i)!}f_2(\mathbf{a})^M,
\end{equation}
where
\begin{eqnarray}
f_2(\mathbf{a})&=&\mathbf{E}[W(\vec\sigma,\vec\tau,c)W(\vec\sigma,\vec\tau,c)]\nonumber\\
&=&2^{-K}\sum_{\vec u,\vec v,\vec u',\vec v'}W(\vec u,\vec v)W(\vec u',\vec v')\prod_{i=1}^{K}a_0^{\delta(u_i=v_i=u'_i=v'_i)}a_1^{\delta(u_i=v_i=u'_i\ne v'_i)}\nonumber\\
&&a_2^{\delta(u_i=v_i=v'_i\ne u'_i)}a_3^{\delta((u_i=v_i)\ne(u'_i=v'_i))}a_4^{\delta(u_i=u'_i=v'_i\ne v_i)}\nonumber\\
&&a_5^{\delta((u_i=u'_i)\ne(v_i=v'_i))}a_6^{\delta((u_i=v'_i)\ne(u'_i=v_i))}a_7^{\delta(u'_i=v'_i=u_i\ne u_i)}
\end{eqnarray}
%M enleve un prime dans l'exposant de a4. Check
$\mathbf{a}$ is a 8-component vector giving the proportion of each type of quadruplets $(\tau_i,\sigma_i,\tau'_i,\sigma'_i)$ ---\,$\vec\tau$ being arbitrarily (but without losing generality) fixed to $(1,\ldots,1)$\,--- as described in the following table:
\begin{center}
\begin{tabular}{ccccccccc}
&$a_0$&$a_1$&$a_2$&$a_3$&$a_4$&$a_5$&$a_6$&$a_7$\\
\hline
$\tau_i$ &+&+&+&+&+&+&+&+\\
$\sigma_i$&+&+&+&+&$-$&$-$&$-$&$-$\\
$\tau'_i$&+&+&$-$&$-$&+&+&$-$&$-$\\
$\sigma'_i$&+&$-$&+&$-$&+&$-$&+&$-$
\end{tabular}
\end{center}
The set $V_N\subset [0,1]^8$ is a simplex 
 specified by:
\begin{equation}\label{simplexN}
\left\{\begin{array}{l}\lfloor N(a_4+a_5+a_6+a_7)\rfloor=\lfloor Nx\rfloor \\
\lfloor N(a_1+a_2+a_5+a_6)\rfloor=\lfloor Nx\rfloor\\
\sum_{i=0}^7 a_i=1\end{array}\right.
\end{equation}
\end{clai}

These three conditions \eqref{simplexN} correspond to the normalization of the proportions and to the enforcement of the conditions $d_{\vec\sigma\vec\tau}=\lfloor Nx\rfloor$, $d_{\vec\sigma'\vec\tau'}= \lfloor Nx\rfloor$. When $N\to\infty$, $V=\bigcap_{N\in\mathbb{N}} V_N$ defines a five-dimensional simplex described by the three hyperplanes:
\begin{equation}\label{simplex}
\left\{\begin{array}{l}a_4+a_5+a_6+a_7=x \\
a_1+a_2+a_5+a_6=x\\
\sum_{i=0}^7 a_i=1\end{array}\right.
\end{equation}

%Note that $V$ can easily be mapped onto the $5$-dimensional cube $[0,1]^5$. %M EIther we use it or we don't put it...
In order to yield an asymptotic estimate of $\mathbf{E}(Z^2)$ we first use the following lemma, which results from a simple approximation of integrals by sums:
\begin{lema}
Let $\psi(\mathbf{a})$ be a real, positive, continuous function of $\mathbf{a}$, and let $V_N$, $V$ be defined as previously.
Then there exists a constant $C_0$ depending on $x$ such that for sufficiently large $N$:
\begin{equation}\label{E3}
\sum_{\mathbf{a}\in V_N\cap \{1/N,2/N,\ldots,1\}^8}\frac{N!}{\prod_{i=0}^{7}(Na_i)!}\psi(\mathbf{a})^N
\leq C_0 N^{3/2} \int_V\ud \mathbf{a}\ e^{N[H_8(\mathbf{a})+\ln \psi(\mathbf{a})]},
\end{equation}
where $H_8(\mathbf{a})=-\sum_{i=1}^8 a_i\ln a_i$.
\end{lema}
A standard Laplace method used on Eq.~(\ref{E3}) with $\psi=2(f_2)^\alpha$ yields:
\begin{clai}
For each $K,x$, define:
\begin{equation}\label{defphi}
\Phi(\mathbf{a})=H_8(\mathbf{a})-\ln 2-2H_2(x)+\alpha\ln f_2(\mathbf{a})-2\alpha\ln f_1(x).
\end{equation}
and let $\mathbf{a}_0\in V$ be the global maximum of $\Phi$ restricted to $V$. Suppose that $\partial_{\mathbf{a}}^2 \Phi(\mathbf{a}_0)$ is definite negative. Then there exists a constant $C_1$ such that, for $N$
sufficiently large,
\begin{equation}\label{cond}
\frac{\mathbf{E}(Z)^2}{\mathbf{E}(Z^2)}\geq C_1\exp(-N\Phi(\mathbf{a}_0)).
\end{equation}
\end{clai}

Obviously $\Phi(\mathbf{a}_0) \ge 0 $ in general. In order to use Lemma \ref{lemmasecond}, one must find  the weights
 $W(\vec u,\vec v)$ in such a way that
$\max_{\mathbf{a}\in V}\Phi(\mathbf{a})= 0$. We first notice that, at the particular point $\mathbf{a}^*$ where the two pairs are uncorrelated with each other,
\begin{equation}
a_0^*=a_3^*=\frac{(1-x)^2}{2},\quad a_1^*=a_2^*=a_4^*=a_7^*=\frac{x(1-x)}{2},\quad
a_5^*=a_6^*=\frac{x^2}{2},
\end{equation}
we have the following properties:
\begin{itemize}
\item $H_8(\mathbf{a}^*)=\ln 2+2H_2(x)$,
\item $\partial_{\mathbf{a}}H_8(\mathbf{a}^*)=0,$ $\partial_{\mathbf{a}}^2 H_8(\mathbf{a}^*)$ definite negative,
\item $f_1(x)^2=f_2(\mathbf{a}^*)$ and hence $\Phi(\mathbf{a}^*)=0$.
\end{itemize}
(Note that the derivatives $\partial_{\mathbf{a}}$ are taken in the simplex $V$).
So the weights must be chosen in such a way that $\mathbf{a}^*$ be
the global maximum of $\Phi$. A necessary condition is that  $\mathbf{a}^*$ be a local
maximum, which entails $\partial_{\mathbf{a}}f_2(\mathbf{a}^*)=0$.

 Using the fact that the number of common values between four vectors $\vec u,\vec v,\vec u',\vec v'\in\{-1,1\}^K$ can be written as:
\begin{equation}
\frac{1}{8}\left(K+\vec u\cdot\vec v+\vec u\cdot\vec u'+\vec u\cdot\vec v'+\vec v\cdot\vec u'+\vec v\cdot\vec v'+ \vec u'\cdot\vec v'+\overrightarrow{u\cdot v}\cdot\overrightarrow{u'\cdot v'}\right)
\end{equation}
we deduce from $\partial_{\mathbf{a}}f_2(\mathbf{a}^*)=0$ the condition:
\begin{equation}\label{eqnln1}
\sum_{\vec u,\vec v}W(\vec u,\vec v)\left\{\begin{array}{l}\vec u\\ \vec v\end{array}\right. (1-x)^{|\overrightarrow{u\cdot v}|}x^{K-|\overrightarrow{u\cdot v}|}=0,
\end{equation}
\begin{eqnarray}
0&=&K(2x-1)^2{\left[\sum_{\vec u,\vec v}W(\vec u,\vec v) (1-x)^{|\overrightarrow{u\cdot v}|}x^{K-|\overrightarrow{u\cdot v}|}  \right]}^2\nonumber\\
&&+{\left[\sum_{\vec u,\vec v}W(\vec u,\vec v) \overrightarrow{u\cdot v}\, (1-x)^{|\overrightarrow{u\cdot v}|}x^{K-|\overrightarrow{u\cdot v}|}\right]}^2\nonumber\\
&&+2(2x-1)\left[\sum_{\vec u,\vec v}W(\vec u,\vec v) \vec u\cdot\vec v\, (1-x)^{|\overrightarrow{u\cdot v}|}x^{K-|\overrightarrow{u\cdot v}|}\right]\nonumber\\
&&\times\left[\sum_{\vec u,\vec v}W(\vec u,\vec v) (1-x)^{|\overrightarrow{u\cdot v}|}x^{K-|\overrightarrow{u\cdot v}|}\right].\label{eqnln2}
\end{eqnarray}

If we suppose that $W$ is invariant under simultaneous and identical permutations of the
$u_i$ or of the $v_i$ (which we must, since the ordering of the variables by the label $i$ is arbitrary), the $K$ components of all vectorial quantities in
Eqs.~(\ref{eqnln1}), (\ref{eqnln2}) should be equal. Then we obtain equivalently:
\begin{align}\label{eqnln1p}
\sum_{\vec u,\vec v}W(\vec u,\vec v)(2|\vec u|-K)\, (1-x)^{|\overrightarrow{u\cdot v}|}x^{K-|\overrightarrow{u\cdot v}|}=0 \quad \textrm{and}\quad \vec u\leftrightarrow \vec v,\\
\label{eqnln2p}
\sum_{\vec u,\vec v}W(\vec u,\vec v)(K(2x-1)+\vec u\cdot\vec v) (1-x)^{|\overrightarrow{u\cdot v}|}x^{K-|\overrightarrow{u\cdot v}|}=0,
\end{align}

We choose the following simple form for $W(\vec u,\vec v)$:
\begin{equation}
W(\vec u,\vec v)=\left\{\begin{array}{ll} 0&\textrm{if}\quad\vec u=(-1,\ldots,-1)\quad\textrm{or}\quad\vec v=(-1,\ldots,-1),\\
\lambda^{|\vec u|+|\vec v|}\nu^{|\overrightarrow{u\cdot v}|} &\textrm{otherwise}.\end{array}\right.
\end{equation}
Although this choice is certainly not optimal, it turns out particularly tractable.
Eqs.~\eqref{eqnln1p} and \eqref{eqnln2p} simplify to:
\begin{equation}\label{lambdanu}
\begin{split}
{[\nu (1-x)]}^{K-1}=&(\lambda^2+1-2\lambda\nu){\left(2\lambda x+\nu (1-x)(1+\lambda^2)\right)}^{K-1}\\
{\left(\nu (1-x)+\lambda x\right)}^{K-1}=&(1-\lambda\nu){\left(2\lambda x+\nu (1-x)(1+\lambda^2)\right)}^{K-1}.
\end{split}
\end{equation}
We found numerically a unique solution $\lambda>0, \nu>0$ to these equations for any value of $K \ge 2$ that we checked.

Fixing $(\lambda,\nu)$ to a solution of \eqref{lambdanu}, we seek the largest value of $\alpha$ such that the
local maximum $\mathbf{a}^*$ is a global maximum, i.e. such that there exists no $\mathbf{a}\in V$ with $\Phi(\mathbf{a})>0$. 
To proceed one needs analytical expressions for $f_1(x)$ and $f_2(\mathbf{a})$. $f_1$ simply reads:
\begin{eqnarray}
f_1(x)&=&2^{-K}{\left((1-x)\nu(1+\lambda^2) +2x\lambda\right)}^K-2\cdot 
2^{-K}{\left(x\lambda+(1-x)\nu\right)}^K\nonumber\\
&&+2^{-K}((1-x)\nu)^K.
\end{eqnarray}
$f_2$ is calculated by Sylvester's formula, but its expression is long and requires preliminar notations. 
%M proposes to change a little bit this section, maybe the new pressentation is a bit simpler, maybe it is 
% just because I am more used to it...
We  index the $16$ possibilities for $(u_i,v_i,u_i',v_i')$ by a number $r \in\{0,\ldots,15\}$ defined as:
\begin{equation}
r=8 \frac{1-u_i}{2}+ 4 \frac{1-v_i}{2} + 2 \frac{1-u_i'}{2} + \frac{1-v_i'}{2} \ .
\end{equation}
For each index $r$, define
\begin{align}
l(r)&=\delta(u_i=1)+ \delta(v_i=1)+\delta(u_i'=1)+ \delta(v_i'=1), \\
n(r)&=\delta(u_i v_i=1)+ \delta(u_i' v_i'=1),
\end{align}
and
\begin{eqnarray}
z_r&=&{\lambda}^{l(r)}  {\nu}^{n(r)} \times\left\{\begin{array}{ll}a_r&\textrm{if}\quad r\leq 7\\a_{15-r}&\textrm{if}\quad  r \ge 8\end{array}\right. \ .
\end{eqnarray}
Also define the four following subsets of $\{0,\ldots,15\}$: $A_0$ is the set of indices $r$ corresponding to quadruplets of the form $(-1,v_i,u_i',v_i')$. $A_0=\{r\in\{0,\ldots,15\}\,|\,u_i=-1\}$. Similarly, $A_1=\{r\,|\,v_i=-1\}$, $A_2=\{r\,|\,u'_i=-1\}$ and $A_3=\{r\,|\,v'_i=-1\}$.

Then  $f_2$ is given by:
\begin{eqnarray}
2^K f_2(\mathbf{a})&=&{\left(\sum_{j=0}^{15}z_j\right)}^{K}-\sum_{k=0}^3{\left(\sum_{j\in A_k}z_j\right)}^{K}+\sum_{0\leq k<k'\leq 3}{\left(\sum_{j\in A_k\cap A_{k'}}z_j\right)}^{K}\nonumber\\
&&-\sum_{0\leq k<k'<k''\leq 3}{\left(\sum_{j\in A_k\cap A_{k'}\cap A_{k''}}z_j\right)}^{K}+{\left(\sum_{j\in A_0\cap A_1\cap A_2 \cap A_3}z_j\right)}^{K}.
\end{eqnarray}

We can now state our lower-bound result:
\begin{lema}\label{th_lb}
Let $\alpha_+\in (0,+\infty]$ be the smallest $\alpha$ such that $\partial^2_\mathbf{a} \Phi(\mathbf{a}^*)$ is not definite negative.
For each $K$ and $x\in (0,1)$, and for all $\alpha\leq\alpha_{LB}(K,x)$, with
\begin{equation}\label{alpha2}
\alpha_{LB}(K,x)=\min\left[\alpha_+,\inf_{\mathbf{a}\in V_+}\frac{\ln
  2+2H_2(x)-H_8(\mathbf{a})}{\ln f_2(\mathbf{a})-2\ln f_1(x)}
\right],
\end{equation}
where $V_+=\{\mathbf{a}\in V\ |\ f_2(\mathbf{a})>f_1^2\left(1/2\right)\}$,
and 
where $(\lambda,\nu)$ is chosen to be a positive solution of (\ref{lambdanu}), the probability
that a random formula
$F_K(N,N\alpha)$ is  $x$-satisfiable is bounded away from $0$ as $N \to\infty$.
\end{lema}
This is a straightforward consequence of the expression (\ref{defphi}) of $\Phi(\mathbf{a})$.

Theorem \ref{fried2} and Lemma \ref{th_lb} immediately imply:
\begin{theo}
For all $\alpha< \alpha_{LB}(K,x)$ defined in Lemma \ref{th_lb},
a random $K$-CNF formula $F_K(N,N\alpha)$
is $x$-satisfiable w.h.p.
\end{theo}

We devised several numerical strategies to evaluate
$\alpha_{LB}(K,x)$. The implementation of Powell's
method on each point of a grid of size $\mathcal{N}^5$
($\mathcal{N}=10,15,20$) on $V$ turned out to be
the most efficient and reliable. The results are given by Fig.~\ref{alpha8} for $K=8$, the smallest $K$ such that the picture given by Conjecture \ref{cluster} is confirmed. We found a clustering phenomenon
for all the values of $K \ge 8$ that we checked. In the following we shall provide a rigorous estimate of $\alpha_{LB}\left(K,\frac{1}{2}\right)$ at large $K$.

\section{Large $K$ analysis}\label{largek}
\subsection{Asymptotics for $x=\frac{1}{2}$}
The main result of this section is contained in the following theorem, which implies Eq.~\eqref{ineq3} in Theorem \ref{nonmonotonic}:
\begin{theo}\label{theolb}
The large $K$ asymptotics of $\alpha_{LB}(K,x)$ at $x=1/2$ is given by:
\begin{equation}
\alpha_{LB}(K,1/2)\sim 2^K\ln 2.
\end{equation}
\end{theo}
The proof primarily relies on the following results:
\begin{clai}\label{lemlambda}
Let $\nu=1$ and $\lambda$ be the unique positive root of:
\begin{equation}\label{eqlambda}
(1-\lambda){(1+\lambda)}^{K-1}-1=0.
\end{equation}
Then $(\lambda,\nu)$ is solution to (\ref{lambdanu}) with $x=\frac{1}{2}$ and one has, at large $K$:
\begin{equation}
\lambda-1 \sim -2^{1-K}.
\end{equation}
\end{clai}
%{\it Proof.} The polynomial \eqref{eqlambda} has value $0$ in $\lambda=0$, and $-1$ in $\lambda=1$. It is increasing for $\lambda\in (0,1-2/K)$ and decreasing for $\lambda >1-2/K$. Therefore that it has a unique positive root $\lambda$, which tends to $1$ as $K$ tends to infinity. The asymptotics at large $K$ follows easily.

\begin{lema}\label{hardlemma1}
Let $x=\frac{1}{2}$. There exist $K_0>0$, $C_1>0$ and $C_2>0$ such that for all $K\geq K_0$, and for all $\mathbf{a}\in V$ s.t. $|\ba-\ba^*|<1/8$,
\begin{equation}\label{2ndmajo}
\left\vert \ln f_2(\mathbf{a})-2\ln f_1(1/2) \right\vert \leq K^2C_1{\vert\mathbf{a}-\mathbf{a}^*\vert}^2 2^{-2K} + C_2 {\vert\mathbf{a}-\mathbf{a}^*\vert}^3 2^{-K}
\end{equation}
\end{lema}

\begin{lema}\label{hardlemma2}
Let $x=\frac{1}{2}$. 
There exist $K_0>0$, $C_0>0$ such that for $K\geq K_0$, for all $\mathbf{a}\in V$,
\begin{equation}\label{1stmajo}
\begin{split}
\left|\ln f_2(\mathbf{a})-2\ln f_1(1/2)\right|\leq  2^{-K}\left[(a_0+a_1+a_4+a_5)^K +(a_0+a_2+a_4+a_6)^K\right. \\
 \left. +(a_0+a_1+a_6+a_7)^K +(a_0+a_2+a_5+a_7)^K\right] + C_0 K2^{-2K}
\end{split}
\end{equation}
\end{lema}

The proofs of these lemmas are defered to sections \ref{proofhardlemma1} and \ref{proofhardlemma2}.

\subsection{Proof of Theorem \ref{theolb}}

We first show that $\partial^2_{\mathbf{a}}\Phi(\mathbf{a}^*)$ is definite negative for all $\alpha<2^K$, when $K$ is sufficiently large. Indeed $\partial^2_{\mathbf{a}}H_8(\mathbf{a}^*)$ is definite negative and its largest eigenvalue is $-4$. Using Lemma \ref{hardlemma1}, for $\mathbf{a}\in V$ close enough to $\mathbf{a}^*$:
\begin{equation}
\Phi(\ba)\leq -2|\ba-\ba^*|^{2}+\alpha C_1{\vert\mathbf{a}
-\mathbf{a}^*\vert}^{2} K^2 2^{-2K} + \alpha C_2 {\vert\mathbf{a}-\mathbf{a}^*\vert}^{3} 2^{-K}.
\end{equation}
Therefore
\begin{equation}
\Phi(\ba)\leq -|\ba-\ba^*|^{2}\quad\textrm{ for }K\textrm{ large enough, }|\ba-\ba^*|<\frac{1}{2C_2}\textrm{ and }\alpha<2^K.
\end{equation}

Using Theorem \ref{th_lb}, we need to find the minimum, for $a\in V_+$, of
\begin{equation}
G(K,{\ba})\equiv \frac{3\ln  2-H_8(\mathbf{a})}{\ln f_2(\mathbf{a})-2\ln f_1(1/2)}.
\end{equation}
We shall show that 
\begin{equation}
\inf_{\ba\in V_{+}} G(K,\ba)\sim 2^K\ln 2.\label{equivG}
\end{equation}
We divide this task in two parts. The first part states that there exists $R>0$ and $K_1$ such that for all $K\geq K_1$, and for all $\ba\in V_+$ such that $|\ba-\ba^*|<R$, $G(K,\ba)>2^K$. This is a consequence of Lemma \ref{hardlemma1}; using the fact that $3\ln 2-H_8(\ba)\geq |\ba-\ba^*|^{2}$ for $\ba$ close enough to $\ba^*$, one obtains:
\begin{equation}
G(K,\ba)\geq \frac{2^K}{C_1 K^2 2^{-K}+C_2 |\ba-\ba^*|}
\end{equation}
which, for $K$ large enough and $\ba$ close enough to $\ba^*$, is greater than $2^K$.

The second part deals with the case where $\ba$ is far from $\ba^*$, i.e. $|\ba-\ba^*|>R$. First we put a bound on the numerator of $G(\ba)$: there exists a constant $C_3>0$ such that for all $\ba\in V$ s.t. $|\ba-\ba^*|>R$, one has $3\ln 2-H_8(\ba)>C_3$.

Looking at Eq.~\eqref{1stmajo}, it is clear that, in order to minimize $G(K,\ba)$, $\ba$ should be `close' to at least one the four hyperplanes defined by
\begin{equation}
\begin{split}
a_0+a_1+a_4+a_5=1, &\qquad
a_0+a_2+a_4+a_6=1,\\
a_0+a_1+a_6+a_7=1, &\qquad
a_0+a_2+a_5+a_7=1.
\end{split}
\end{equation}
More precisely, we say for instance that $\ba$ is \emph{close to} the first hyperplane defined above iff
\begin{equation}
a_0+a_1+a_4+a_5>1-K^{-1/2}
\end{equation}
Now suppose that $\ba$ is \emph{not} close to that hyperplane. Then the corresponding term goes to $0$:
\begin{equation}
(a_0+a_1+a_4+a_5)^K\leq \left(1-K^{-1/2}\right)^{K}\sim \exp(-\sqrt{K})\quad\textrm{as }K\to \infty.
\end{equation}

We classify all possible cases according to the number of hyperplanes $\ba\in V_+$ is close to:
\begin{itemize}
\item 
$\mathbf{a}$ is close to none of the hyperplanes. Then
\begin{equation}
G(K,\ba)\geq \frac{2^K C_3}{4\exp(-\sqrt{K})+C_0K2^{-K}}>2^K\qquad\textrm{for }K\textrm{ large enough.}
\end{equation}
\item 
$\mathbf{a}$ is close to one hyperplane only, e.g. the first hyperplane $a_0+a_1+a_4+a_5=1$ (the other hyperplanes are treated equivalently). As $\sum_{i=0}^{7}a_i=0$, one has
\begin{equation}
a_2<K^{-1/2},\quad a_3<K^{-1/2},\quad a_6<K^{-1/2},\quad a_7<K^{-1/2}.
\end{equation}
This implies $H_{8}(\ba)<2\ln 2 + 2\ln K/\sqrt{K}$, and we get:
\begin{equation}
G(K,\ba)\geq \frac{2^K [\ln 2-2\ln K/\sqrt{K}]}{1+C_0K 2^{-K}+3\e^{-\sqrt{K}}}\geq 2^K (\ln 2) \left[1-3\ln K/\sqrt{K}\right]
\end{equation}
for sufficiently large $K$.
\item
$\ba$ is close to two hyperplanes. It is easy to check that these hyperplanes must be either the first and the fourth ones, or the second and the third ones. In the first case we have $a_0+a_5>1-3/\sqrt{K}$ and in the second case $a_0+a_6>1-3/\sqrt{K}$. Both cases imply: $H_{8}(\ba)<\ln 2 + 3\ln K/\sqrt{K}$. One thus obtains:
\begin{equation}
G(K,\ba)\geq \frac{2^K [2\ln 2-3\ln K/\sqrt{K}]}{2+C_0 K2^{-K}+2\e^{-\sqrt{K}}}\geq 2^K (\ln 2) \left[1-3\ln K/\sqrt{K}\right].
\end{equation}
\item One can check that $\ba$ cannot be close to more than two hyperplanes.
\end{itemize}
To sum up, we have proved that for $K$ large enough, for all $\ba\in V_+$,
\begin{equation}
G(K,\ba)\geq 2^K (\ln 2) \left[1-3\ln K/\sqrt{K}\right],
\end{equation}
Clearly, $\alpha_{LB}(K,1/2)=\inf_{\ba\in V_+} G(K,\ba)<\alpha_{UB}(K,1/2)$. Since from Theorem \ref{thUB} we know that
$\alpha_{UB}(K,1/2)\sim 2^K\ln 2$, this proves Eq.~\eqref{equivG}.

\subsection{Proof of Lemma \ref{hardlemma1}}\label{proofhardlemma1}

Let $x=\frac{1}{2}$ and choose $\nu=1$ and $\lambda$ the unique positive root of Eq.~\eqref{eqlambda}. Let $\epsilon_i=a_i-1/8$, and $\be=(\epsilon_0,\ldots,\epsilon_7)$. We expand $f_2(\ba)$ in series of $\be$. The zeroth order term is $f_2(1/8,\ldots,1/8)=f_1^2(1/2)$. The first order term vanishes. We thus get:
\begin{equation}\label{f2dev}
f_2(\ba)=f_1^{2}(1/2)+B_0-B_1+B_2-B_3+B_4,
\end{equation}
with
{\allowdisplaybreaks
\begin{eqnarray}
B_0&=&\sum_{q=2}^K\binom{K}{q}{\left(\frac{1}{2}\sum_{i=0}^7 p_i(\lambda)\epsilon_i\right)}^q {\left[\frac{1+\lambda}{2}\right]}^{4(K-q)},\\ 
B_1&=& 2^{-K}\sum_{a=1}^4\sum_{q=2}^K \binom{K}{q} {\left[\sum_{i=0}^7 \left(\lambda^{\ell_{ai}}-1\right)\epsilon_i\right]}^q {\left[\frac{1+\lambda}{2}\right]}^{3(K-q)},\\
B_2&=& 2^{-2K}\sum_{a=1}^6\sum_{q=2}^K \binom{K}{q} [2r_a(\lambda,\be)]^q {\left[\frac{1+\lambda}{2}\right]}^{2(K-q)},\\
B_3&=& 2^{-3K}\sum_{a=1}^4\sum_{q=2}^K \binom{K}{q} [4s_a(\lambda,\be)]^q {\left[\frac{1+\lambda}{2}\right]}^{K-q},\\
B_4&=& 2^{-4K} \sum_{k=2}^K (8\epsilon_0)^q.
\end{eqnarray}}
In $B_0$, $p_i(\lambda)=\lambda^{l(i)}+\lambda^{l(15-i)}-2-4(\lambda-1)$. We have used the fact that $\sum_{i=0}^7 \epsilon_i=0$. Using $l(i)+l(15-i)=4$, one obtains $|p_i(\lambda)|\leq 11(\lambda-1)^{2}\leq 11 \cdot 2^{4-2K}$, since $|\lambda-1|\leq 2^{2-K}$ for $K$ large enough, by virtue of Lemma \ref{lemlambda}.

In $B_1$, we have used again $\sum_{i=0}^7 \epsilon_i=0$. $\ell_{ai}$ is either $l(i)$ or $l(15-i)$, depending on $a$. In both cases $|\lambda^{\ell_{ai}}-1|\leq 4|\lambda-1|\leq 2^{4-K}$. In $B_2$ and $B_3$, the expressions of $r_a(\lambda,\be)$ and $s_{a}(\lambda,\be)$ are given by:
\begin{equation}
\begin{split}
r_1=\epsilon_0+\lambda(\epsilon_1+\epsilon_2)+\lambda^{2}\epsilon_3, &\qquad r_2=\epsilon_0+\lambda(\epsilon_1+\epsilon_4)+\lambda^{2}\epsilon_5,\\
r_3=\epsilon_0+\lambda(\epsilon_2+\epsilon_4)+\lambda^{2}\epsilon_6, &\qquad r_4
=\epsilon_0+\lambda(\epsilon_1+\epsilon_7)+\lambda^{2}\epsilon_6,\\
r_5=\epsilon_0+\lambda(\epsilon_2+\epsilon_7)+\lambda^{2}\epsilon_5, &\qquad r_6
=\epsilon_0+\lambda(\epsilon_4+\epsilon_7)+\lambda^{2}\epsilon_3,
\end{split}
\end{equation}
\begin{equation}
s_1=\epsilon_0+\lambda \epsilon_1,\qquad
s_2=\epsilon_0+\lambda \epsilon_2,  \qquad
s_3=\epsilon_0+\lambda \epsilon_4,\qquad 
s_4=\epsilon_0+\lambda \epsilon_7.\qquad
\end{equation}

In order to prove Lemma \ref{hardlemma1} we will use the following fact:
\begin{clai}
Let $y$ be a real variable such that $|y|\leq 1$. Then 
\begin{equation}
\left|\sum_{k=2}^K \binom{K}{k} y^k\right|\leq \frac{K(K-1)}{2}y^{2}+2^K |y|^{3}.
\end{equation}
\end{clai}

%Recall that $\lambda<1$. Let us check that |2r_{a}|\leq 1. For instance, for $a=1$, one has:
%\begin{equation}
%0<(\frac{(1+\lambda)^{2}}{4}+r_1=a_0+\lambda(a_1+a_2)\lambda^{2}a_3<

One has $|2r_{a}|\leq 8|\be|$, $|4s_{a}|\leq 8|\be|$, and $|8\epsilon_0|\leq 8|\be|$. Therefore, for $|\be|<1/8$, one can write:
\begin{eqnarray}
|B_0|&\leq & \frac{K(K-1)}{2}(11\cdot 2^6)^{2} 2^{-4K} |\be|^{2} + (11\cdot 2^6)^{3} 2^{-5K} |\be|^{3}\label{majo0}\\
|B_1|&\leq & 4\frac{K(K-1)}{2}2^{14} 2^{-3K} |\be|^{2} + 2^{21} 2^{-3K} |\be|^{3}\label{majo1}\\
|B_i|&\leq & \binom{4}{i}\frac{K(K-1)}{2} 2^6 2^{-iK} |\be|^{2} + 2^9 2^{-(i-1)K} |\be|^{3}\quad\textrm{for }2\leq i\leq 4.
\end{eqnarray}

Observe that
\begin{equation}\label{asympf}
f_1(1/2)={\left[{\left(\frac{1+\lambda}{2}\right)}^K-2^{-K}\right]}^{2}=1+O(K2^{-K})
\end{equation}
and that for $K$ large enough,
\begin{equation}
\left|\ln\frac{f_2(\ba)}{f_1^2(1/2)}\right|\leq \frac{2}{f_1(1/2)^2}\sum_{i=0}^{4}|B_i|,
\end{equation}
which proves Lemma \ref{hardlemma1}.

\subsection{Proof of Lemma \ref{hardlemma2}}\label{proofhardlemma2}
Note that the bounds on $B_0$ and $B_1$ \eqref{majo0}, \eqref{majo1} remain valid for any $\be$. Therefore $B_0=O(2^{-2K})$ and $B_1=O(2^{-2K})$ uniformly. We bound $B_3$ by observing that:
\begin{equation}
\begin{split}
B_3=&2^{-K}\left[(a_0+\lambda a_1)^K+(a_0+\lambda a_2)^K+(a_0+\lambda a_4)^K+(a_0+\lambda a_7)^K\right]\\
&-2^{-3K}\sum_{a=1}^4{\left[\frac{1+\lambda}{2}\right]}^K\left[1+K \left(\frac{8s_a(\lambda,\be)}{1+\lambda}\right)\right].
\end{split}
\end{equation}
Since $(a_0+\lambda a_1)\leq a_0+a_1\leq 1/2$ and likewise for the three other terms, one has $B_3=O(2^{-2K})$ uniformly in $\ba$. A similar argument yields $B_4=O(2^{-2K})$. There remains $B_2$, which we write as:
\begin{equation}
\begin{split}
B_2=&2^{-K}\sum_{0\leq k<k'\leq 3}{\left(\sum_{j\in A_k\cap A_{k'}}z_j\right)}^{K}\\
&-2^{-2K}\sum_{a=1}^6{\left[\frac{1+\lambda}{2}\right]}^{2K}\left[1+K \left(\frac{8r_a(\lambda,\be)}{(1+\lambda)^2}\right)\right]
\end{split}
\end{equation}
The second term of the sum is $O(K2^{-2K})$. The first term is made of six contributions. Two of them, namely $2^{-K}(a_0+\lambda(a_1+a_2)+\lambda^2 a_3)$ and $2^{-K}(a_0+\lambda(a_4+a_7)+\lambda^2 a_3)$, are $O(2^{-2K})$, because of the condition on distances. Among the four remaining contributions, we show how to deal with one of them, the others being handled similarly. This contribution can be written as:
\begin{equation}
(a_0+\lambda(a_1+a_4)+\lambda^2 a_5)^K=(a_0+a_1+a_4+a_5)^K {\left(1+\frac{(\lambda-1)(a_1+a_4)+(\lambda^2-1)a_5}{a_0+a_1+a_4+a_5}\right)}^K.
\end{equation}
We distinguish two cases. Either $a_0+a_1+a_4+a_5\leq 1/2$, and we get trivially:
\begin{equation}
(a_0+\lambda(a_1+a_4)+\lambda^2 a_5)^K-(a_0+a_1+a_4+a_5)^K=O(2^{-K}),
\end{equation}
since both terms are $O(2^{-K})$; or $a_0+a_1+a_4+a_5\geq 1/2$, and then:
\begin{equation}
\begin{split}
\left|(a_0+\lambda(a_1+a_4)+\lambda^2 a_5)^K-(a_0+a_1+a_4+a_5)^K\right|\leq\\
\qquad\left|{\left(1+\frac{(\lambda-1)(a_1+a_4)+(\lambda^2-1)a_5}{a_0+a_1+a_4+a_5}\right)}^K-1\right|=O(K2^{-K}).
\end{split}
\end{equation}
Using again Eq.~\eqref{asympf} finishes the proof of Lemma \ref{hardlemma2}.\hfill $\Box$

\subsection{Heuristics for arbitrary $x$}

For arbitrary $x$, the function to minimize in (\ref{alpha2}) is hard to study analytically.
Here we present what we believe to be the correct asymptotic expansion of $\alpha_{LB}(K,x)$ at large $K$. Hopefully this temptative analysis could be used as a starting point towards a rigorous analytical treatment for any $x$.

A careful look at the numerics suggests the following Ansatz on the position of the global maximum, at large $K$:
\begin{equation}
\label{ans}
\begin{split}
&a_0=1-x+o(1),\quad a_6=x+o(1)\\
&a_i=o(1)\quad\text{for }i\neq 0,6.
\end{split}
\end{equation}
A second, symmetric, maximum also exists around $a_0=1-x$, $a_5=x$. Plugging this locus into Eq.~\eqref{alpha2} leads to the following conjecture:
\begin{conjec}
For all $x\in(0,1]$, the asymptotics of $\alpha_{LB}(x)$ is given by:
\begin{equation}
\lim_{K\to \infty}2^{-K}\alpha_{LB}(K,x)=\frac{\ln 2+H(x)}{2},
\end{equation}
and the limit is uniform on any closed sub-interval of $(0,1]$.
\end{conjec}
This conjecture is consistent with both our numerical simulations and our result at $x=\frac{1}{2}$.

\section{Proof of Theorem \ref{fried2}} \label{prooffriedgut}

Starting with  the sharpness criterion for monotone properties of the hypercube given by E. Friedgut  and J. Bourgain, we will prove Theorem \ref{fried2}  by using techniques and tools developped  by N. Creignou and H. Daud\'e for proving the sharpness of monotone properties in random CSPs. \\

First we make precise some notations for this study on random $K$-CNF formula over $N$ Boolean variables   $  \{x_1, \ldots,  x_ N\}  .$ A $K$-clause $C$ is given in disjunctive form:   $C=x_1^{\varepsilon_1}\vee  \ldots \vee x_K^{\varepsilon_K}$ where $\varepsilon_i \in \{0,1\}$ ($x_i^0$ is the positive literal $x_i$  and   $x_i^1$ is the negative one  $\overline {x_i}$).    A $K$-CNF formula $F$ is a finite conjunction of $K$-clauses,    $\Omega(F)$ will denote  the set of distinct variables occurring in $F$,  $\Omega(F)\subset  \{x_1, \ldots,  x_ N\}  $.  In this Boolean framework,  $S(F)$ the   set of satisfying assignments to $F$, becomes a subset of $\{0,1\}^N$.\\

Now,  let us recall how  a slight change of our probability measure on formul{\ae}  gives a convenient product probability space for studying $x$-satisfiability.

\subsection{$x$-unxatisfiability as a monotone property}

 In our case the number of clauses  in a random formula   $F_K(N, N\alpha)$ is fixed to $M=N\alpha$. We define another kind of random  formula $G_K(N, N\alpha)$ by allowing each of the ${\cN} = 2^K {N\choose K}$ possible clauses to be present  with probability $p= \alpha N / {\cN}$. Then, assigning $1$ to each clause if it is present and $0$ otherwise, the hypercube $\{0,1\}^{\cN}$ stands for the set of all possible formul{\ae}, endowed with the so-called product measure $\mu _p$, where $p$ is the probability for $1$, and $1-p$ for $0$. 

\noindent More generally, let $\cN$ be a positive integer, a property $Y\subset\{0,1\}^{\cN}$ is called monotone if , for any $y, y' \in  \{0,1\}^{\cN}$, $y\leq y'$ and $y \in Y$ implies $y' \in Y$.  In that case $\mu_p(  y \in Y )$ is an increasing function of $p\in [0,1]$ where
$$\mu_p(y_1,\cdots, y_{\cN})=p^{\vert y \vert } \cdot (1-p)^{{\cN}- \vert y \vert} \hbox { where } \vert y \vert = {\sharp \{1\leq i \leq {\cN} \ / \  y_i =1\}}.$$
 For any  non trivial $Y$   we can define  for every $\beta \in ]0,1[$  the unique $p_{\beta}\in ]0,1[$ such that:
  $$\mu_{p_{\beta}}(y \in Y ) = \beta.$$

\noindent In our case $Y$ will be the property of being $x$-unsatisfiable.
If we put: 
\begin{equation}\label{subsetD}
{\cD} =\Bigl \{ (\vec{\sigma}, \vec{\tau}) \in \{0,1\} ^N \times \{0,1\} ^N \quad s. t.\quad  d_{\vec{\sigma} \vec{\tau}} \in [Nx - \varepsilon(N), Nx + \varepsilon(N) ] \Bigr \}
\end{equation}
then $x$-unsatisfiability can be read:
$$F \in Y \Longleftrightarrow  S(F) \times S(F) \quad \cap \quad {\cD} = \emptyset.$$
Observe that the  number  of clauses in $G_K(N, N\alpha)$  is distributed as a binomial  law $\mathrm{Bin} ( {\cN}, p=\alpha N / {\cN})$ peaked around its expected value  $p\cdot {\cN}= \alpha N $. Therefore, from well known results  on monotone property of the hypercube, \cite[page 21 and Corollary 1.16 page 19]{JansonLR-99}, our  Theorem \ref{fried2} is equivalent to the following result, which establishes the sharpness of  the monotone property $Y$ under $\mu_p$.
 
 \begin{theo}\label{sharpmodel2}
For each $K\geq 3$ and $x, 0<x<1$, there exists a sequence $\alpha_N(K,x)$ such that for all $\eta >0$:
\begin {equation}
\lim_{N\to \infty} \mu_p( F \mbox { is } x-unsatisfiable) = \left\{ \begin{array} {rl} 1 \mbox { if } p\cdot {\cN}=(1-\eta ) \alpha_N(K,x)N, \\ 0 \mbox{ if }  p\cdot {\cN}=(1+\eta ) \alpha_N(K,x)N. \end{array} \right.
\end {equation}
\end{theo} 

This theorem will be proved using general results on monotone properties of the hypercube. We state these results below without proof.

 \subsection{General tools}
The main tool used to prove the existence of a sharp threshold will be a sharpness criterion stemming from Bourgain's result  \cite{Friedgut} and  from a  remark by Friedgut on the possibility to strengthen his criterion \cite[Remark following Theorem 2.2]{Friedgut-05}. Thus, a slight strengthening  of Bourgain's proof in the appendix of  \cite{Friedgut} combined with  an observation made in \cite[Theorem 2.3, page 130]{CreignouD-03} gives the following sharpness criterion: 
 
 \begin{theo}\label{sharpcriterion}
 
 Let $Y_{\cN}\subset \{0,1\}^{{\cN}}$ be a sequence of  monotone properties,  then $Y$ has a sharp threshold as soon as there exists a sequence $T_{\cN}$ with $T_{\cN}\supset Y_{\cN}$ such that for any $\beta \in ]0,1[$ and every $D\geq 1$ the three following conditions are satisfied: 
\begin{equation}\label{cond0}
 p_{\beta}=o(1),
\end{equation}
\begin {equation}\label{cond1}
\mu_{p_{\beta}}  (y \ \mathrm {s.t.} \    \exists\,  z \in T, \  z\subset y, \  \vert z \vert \leq D ) = o(1),
 \end{equation}
\begin {equation}\label{cond2}
 \forall \, z_0 \notin T, \,  \vert z_0 \vert \leq D \quad 
\mu_{p_{\beta}}  (y \in Y ,\ y\setminus z_0 \notin Y  \quad \vert  \quad y \supset z_0 \ ) = o(1).
\end{equation} 
 \end{theo} 
We end this subsection by recalling two general results on monotone properties defined on
finite sets, established in \cite{CreignouD-04}.

\begin{lema} \label{perco1}\cite[Lemma A.1, page 236] {CreignouD-04}

Let $U=\{1,\ldots, {\cN}\}$  be  partitioned  into two sets $U'$
and $U''$ with $\#U'={\cN}', \#U''={\cN}''$ and ${\cN}={\cN}'+{\cN}''$.  For any $u\subset U$ let us denote $u'=u\cap U'$ and $u''= u\cap U''. $
Let $Y\subset \{0,1\}^{{\cN}}$ be a monotone property.
        For any element  $u$,   let  ${\cA}(u)$ be the set of
     elements from  $U'$ that are essential for property $Y$ at $u$:
$   {\cA}(u) = \left\{ i\in U' \hbox { s.t. }  u \cup  \{i\} \in Y
\right\}.$ Then,  for any $a>0$  the following holds
$$ \mu_{p} (u\in Y, u''\not\in Y) \le \frac{1}{(1-p)^{{\cN}'}}\cdot\mu_{p}
(u\not\in Y, \#{\cA}(u)\ge a) + \frac{a\cdot p}{(1-p)^{{\cN}'}} .$$
\end{lema}

For the second  result we consider a sequence of  monotone properties $Y_{\cN}\subset \{0,1\}^{{\cN}}$. For any fixed $u\in \{0,1\}^{{\cN}}$,   ${\cB}_j (u)$ will be  the set of collections of $j$ elements  such that one can reach property $Y$ from $u$ by adding  this collection, thus $\#{\cB}_j (u)\le {{\cN} \choose j}$. 

\begin{lema} \label{perco2}\cite[Lemma A.2, page 237] {CreignouD-04}
Let $Y_{\cN}\subset \{0,1\}^{{\cN}}$ be a sequence of  monotone properties.  For any
integer $j\geq 1$, for any $b>0$ and as soon as
${\cN}\cdot p$ tends to infinity, the
     following estimate holds
$$ \mu_{p} \Bigl (u\not\in Y, \#{\cB}_j (u)\ge b\cdot {{\cN} \choose
j}\Bigr) =o(1), $$       
$${\cB}_j (u)=\left\{
\{i_1,\ldots, i_j\}, 1\leq i_1<\ldots< i_j\leq {\cN},   \hbox{ such that } u\cup \{ i_1,\ldots,
i_j\}\in Y\right\}.$$
\end{lema}

\subsection{Proof of Theorem \ref {sharpmodel2} (main steps)}

As usual, the first two conditions $\eqref{cond0}$ and $\eqref{cond1}$ are easy to verify for the $x$-unsatisfiability property.
For the first one we have:  
$$\mu_{p}(F \textrm{ is }x\textrm{-satisfiable})\leq \mu_{p}(F \textrm{ is satisfiable} )\leq 2^N (1-p)^{ N \choose K }.$$
This shows that  $\displaystyle p_{\beta} \leq {N \ln (2)-\ln(1-\beta) \over {N\choose K}}$ , thus  for $x$-unsatisfiability we get:
\begin{equation}\label{cond0OK}
\forall \beta \in ]0,1[ \quad  p_{\beta}(N)= O(N^{1-K}).
\end {equation}

For the second condition, let  $H(F)$ be  the $K$-uniform hypergraph associated to a formula $F$: its vertices are the  $\Omega(F)$ variables occurring in $F$,  each index set  of a clause $C$ in  $F$ corresponds to an hyperedge.  Let us recall, see  
 \cite{KL-02}, that a $K$-uniform connected hypergraph with $v$ vertices and $w$
edges is called a \emph{hypertree} when $(K-1) w-v=-1$; it is
said to be \emph{unicyclic} when   $(K-1) w-v=0$, and \emph{complex} when $(K-1) w-v \geq 1$. Let $T$ be the set of formul{\ae} $F$  such that  $H(F)$ has at least one  complex component. We will rule out $\eqref{cond1}$ (and also $\eqref{cond2}$)  by using  the following  result on non complex formul{\ae}, the proof of which is deferred to the next subsection:

\begin{lema}\label{simple} Let $K\geq 3$. If $G$ is a $K$-CNF-formula on $v$ variables whose associated hypergraph is an hypertree or unicyclic then for all integer $d\in\{0, \ldots, v\} $ there exits $(\vec \sigma,\vec \tau) \in S(G) \times S(G) $ such that $d_{\vec \sigma \vec \tau}=d.$
\end{lema}

\noindent In particular, this result shows that  any $x$-unsatisfiable formula has at least one complex component, i. e. $T\supset Y. $ Then observe that there is $O(N^{(K-1)  s -1})$ distinct complex components  of size $s$ with $N$ vertices. Thus we get for all $p:$
$\displaystyle \mu_{p}  (F \ \mathrm {s.t.} \    \exists\,  G \in T, \  G \subset F, \  \vert G \vert \leq D ) \leq \sum_{s\leq D} O(N^{(K-1)s -1}) \cdot  p^s, $
and  $\eqref{cond1}$ follows from  $\eqref{cond0OK}$\\

In order  to prove $\eqref{cond2}$, let us  introduce some tools inspired of  \cite{CreignouD-04}.

\noindent For each positive integer $t$ and 
$\Delta=(\Delta_1,\ldots ,\Delta_t)\in \{0,1\}^t$, a $\Delta$-assignment
is an assignment for which the $t$ first values of the variables are equal to $\Delta_1,\ldots ,\Delta_t$. 
  Then $S_{\Delta}(F)$ will denote the set of satisfying $\Delta$-assignments to  $F$: $S_{\Delta}(F) \subset S(F) \subset \{0,1\}^N$.
  
  \noindent For any pair of $t$-tuples $(\Delta, \Delta ' ) \in \{0,1\}^t \times \{0,1\}^t$ we define $Y^{\Delta, \Delta'}$:  
  $$ F \in Y^{\Delta, \Delta'} \Longleftrightarrow  S_{\Delta}(F) \times S_{\Delta '}(F) \quad \cap \quad {\cD}_x = \emptyset. $$
Observe that  $Y^{\Delta, \Delta'}$ is a monotone property containing $Y$.

\medskip 
Now we come back to $\eqref{cond2}$ with  $F_0 \notin T$, so that the hypergraph associated to the booster formula $F_0$ has no complex  components.  $S(F_0) \not= \emptyset$ and  w.l.o.g. we  can suppose that $\Omega(F_0)=\{1,\ldots, t\}$. Then,  for $ F \in Y$ such that   $F \supset F_0$ with   $F\setminus F_0 \notin Y$, let $F''$  denote the largest subformula of $F$ such that $\Omega(F'') \cap \{1,\ldots,t\} =\emptyset$.  We have the two following claims whose proof is postponed to  the next subsection.

 \begin{clai}\label{claim1}  For any $(\Delta, \Delta ' ) \in S(F_0) \times S(F_0), \quad   F \setminus F_0 \in Y^{\Delta, \Delta'}$.
 \end{clai}
 
 \begin{clai}\label{claim2}  There exits $(\Delta, \Delta ' ) \in S(F_0) \times S(F_0)$  such that  $   F '' \notin Y^{\Delta, \Delta'}$.
 \end{clai}
 Thus $\eqref{cond2}$ is proved as soon as  for  any $ \beta  \in]0,1[$ and $(\Delta, \Delta ' ) \in  \{0,1\}^{t}\times  \{0,1\}^{t}$: 
  \begin{equation}\label{fromCD2a}
  \mu_{p_{\beta}}  (\  F\setminus F_0 \in Y^{\Delta, \Delta'},    F '' \notin Y^{\Delta, \Delta'} \quad \vert  \quad F \supset F_0 \  ) = o(1).
\end{equation}
\noindent The two first events in the R.H.S. of    (\ref{fromCD2a}) do not depend on the set of clauses in $F_0$ thus by independence under the product measure and  recalling that $Y^{\Delta, \Delta'}$ is a monotone property we are led to prove that:
 $$ \mu_{p_{\beta}}  (\  F \in Y^{\Delta, \Delta'},    F '' \notin Y^{\Delta, \Delta'}\  ) = o(1).$$      
   From $\eqref{cond0OK}$  we know that  
$p_{\beta}(N)=O(N^{1-K})$. Let   ${\cN}'=\Theta(N^{K-1})$ be the number of clauses having at least one variable in $\{1,\ldots,t\}$,  then Lemma \ref{perco1}, applied to the monotone property $Y^{\Delta, \Delta'}$, shows that the above assertion is true as soon as we are able to prove that  for all $\gamma >0$:

\begin{equation}\label{fromCD2b}
  \mu_{p_{\beta}}  (\  F \notin Y^{\Delta, \Delta'},   \# {\cA}_{\Delta, \Delta'}(F) \geq \gamma \cdot  N^{K-1} \ ) =o(1).
\end{equation} 

\noindent where   ${\cA}_{\Delta, \Delta'}(F)$ is the set of $K$-clauses $C$ on $N$ variables having at least one variable in $ \{x_1, \ldots ,x_t \} $ and such that $F \wedge C \in  Y^{\Delta, \Delta'}$.

Then  let ${\cB}_{\Delta, \Delta'}(F)$ be  the set of collections of $(K-1)$  $K$-clauses
$\{C_1, \ldots , C_{K-1}\}$  such that $F \wedge C_1\wedge \ldots  \wedge C_{K-1} \in  Y^{\Delta, \Delta'}$. From lemma  \ref{perco1} we deduce  that  (\ref{fromCD2b}) is true as soon as the following result is proved:  

\begin{lema}\label{comblemma} For all $ t,  K\geq 3, \gamma>0$ and $(\Delta, \Delta ' ) \in \{0,1\}^t \times \{0,1\}^t$, there exits $\theta >0$ such that  for all $N$, the following holds:
\begin{equation}\label{fromCD3}
 \# {\cA}_{\Delta, \Delta'}(F) \geq \gamma \cdot  N^{K-1}  \Longrightarrow     \# {\cB}_{\Delta, \Delta'}(F) \geq \theta \cdot  N^{K \cdot (K-1)}.
  \end{equation}

\end{lema} 
 Again the proof of this last result is deferred to the next subsection that furnishes a  detailed and complete proof of Theorem \ref {sharpmodel2}.
\subsection{Detailed proofs}\label{Proofs}

\subsubsection{Lemma \ref{simple}} 

\begin{proof}
When $G$ has a leaf-clause, that is a clause $C=x_1^{\varepsilon_1}\vee  \ldots \vee x_K^{\varepsilon_K}$ having only one variable, say $x_1$,  in common with $G \setminus C$,  the assertion can be proved by   induction on the number of clauses in $G$. Indeed from a pair of satisfying assignments $(\vec \sigma,\vec \tau) \in S(G \setminus C) \times S( G\setminus C)$ with $d_{\vec \sigma \vec \tau}=d $ and  a pair of satisfying assignments at distance $d' \in\{0, \ldots, K-1\}$ for $C'=x_2^{\varepsilon_2}\vee  \ldots \vee x_K^{\varepsilon_K}$,  one  gets a pair of satisfying assignments at  distance  $d+d'$.  But $C'$ is a $K-1$-clause, thus for any $d' \in \{0, \ldots, K-1\}$  $C'$  has  a pair of satisfying assignments at distance $d'$.\\

When any  $K$-clause $C_i$  of $G=C_1\wedge \ldots\wedge C_l$ has exactly two variables in common with $G\setminus C_i$ then  we can write $C_1=x_1^{\mu_1}\vee x_2^{\nu_2} \vee C'_1, C_2=x_2^{\mu_2}\vee x_3^{\nu_3} \vee C'_2, \ldots,   C_l=x_l^{\mu_l}\vee x_1^{\nu_1} \vee C'_l$ where the $C'_j$ are $(K-2)$-clauses. A variable  in $C'_j$ occurs exactly once in formula $G$ and the   set  of variables  in these $C'_j$ is equal to $\{ x_{l+1}, \ldots,x_v\}$. In particular this set is disjoint from the set of variables of the $2$-CNF  formula  $(x_1^{\mu_1}\vee x_2^{\nu_2}) \wedge (x_2^{\mu_2}\vee x_3^{\nu_3}) \wedge \ldots   \wedge(x_l^{\mu_l}\vee x_1^{\nu_1})$. First observe that this $2$-CNF cyclic formula  has always a  satisfying assignment $(\sigma_{1}, \ldots, \sigma_{l})$ and  together with any truth value for the $(x_j, j >l)$ it gives a  satisfying assignment for  $G$. Thus, for $G$,   one gets  a 
pair of satisfying assignments at distance $d$ for any $d\leq v-l$. Second, as $\Omega(C'_j) \cap \Omega(C'_k)=\emptyset$ when $j \not = k$  a satisfying assignment $\sigma_{l+1}, \ldots, \sigma_{v}$ can easily be found for $ C'_1\wedge \ldots \wedge    C'_l$.  Together with any truth values  of the $(x_i, i\leq l) $  it gives  a satisfying assignment for  $G$. Then,  from the satisfying assignment $(\sigma_{1}, \ldots, \sigma_{l},1-\sigma_{l+1}, \ldots, 1-\sigma_{v} )$   one gets, for any $d\geq v-l$,  a pair of satisfying assignments at distance $d$.\end{proof}

\subsubsection{Claims \ref{claim1} and \ref{claim2}} 

\begin {proof}
Observe that  any SAT-$x$-pair $(\vec \sigma,\vec \tau)$ for $F\setminus F_0$ with $(\sigma_1, \ldots, \sigma_t) \in S(F_0)$ and   $(\tau_1, \ldots, \tau_t) \in S(F_0)$ is also a SAT-$x$-pair for $F$. This proves the first claim by contradiction.\\

For the second claim,   $F\setminus F_0 \notin Y$  so there exits a SAT-$x$-pair $(\vec \sigma,\vec \tau) \in S(F\setminus F_0) \times S(F\setminus F_0) $. By construction, the set of satisfying assignment of $F''$ does not depend on the first $t$ coordinates.  Let $d_t$ be the Hamming distance between $(\sigma_1,\ldots \sigma_t)$ and $(\tau_1,\ldots \tau_t)$. We know  that all components of the hypergraph associated to formula $F_0$ are simple and lemma $\eqref{simple}$ shows that there exits $(\sigma'_1,\ldots \sigma'_t) \in S(F_0)$ and $(\tau'_1,\ldots \tau'_t) \in S(F_0) $ such that $d_{\vec \sigma'\vec \tau'}=d_t$. Hence $(\sigma'_1,\ldots \sigma'_t, \sigma_{t+1}, \ldots, \sigma_N)$ and $(\tau'_1,\ldots \tau'_t, \tau_{t+1}, \ldots, \tau_N)$ form now a SAT-$x$-pair for $F''$, thus proving the second claim.\end{proof}

 \subsubsection{ Lemma \ref{comblemma}}
 
 \begin{proof}
  In \cite{ErdosS-82},  Erd\"os and Simonovits proved
that  any sufficiently dense uniform hypergraph always contains specific
 subhypergraphs. In particular they considered
  a generalization of the complete
bipartite graph specified by two integers $h\ge 2$ and $m\ge 1$.
 Let us denote by   $K_{h}(m)$   the $h$-uniform hypergraph
  with $h\cdot m$ vertices partitioned into $h$ classes $V_1, \cdots, V_h$
  with $\#V_i =m$ and whose
hyperedges  are those $h$-tuples, which have exactly one vertex in
each $V_i$.   Thus  $K_{h}(m)$  has $m^h$ hyperedges,   for $h=2$ it
is a complete bipartite graph $K(m,m)$.

 For proving  Lemma \ref{comblemma}, we  need  a small variation on a result
 of Erd\"os and Simonovits which differs only in that it deals
 with ordered $h$-tuples as opposed to sets of size $h$. More precisely,
 let us consider hypergraphs on $n$ vertices, say $\{x_1,\ldots, x_n\}$,
 we will say that two disjoint subsets of
vertices $A$ and $B$ verify $A<B$ if for all $x_i$ in $A$ and all
$x_j$ in $B$ we have $i<j$. Let $H$ be an $h$-uniform hypergraph
with vertex set $\{x_1,\ldots, x_n\}$, then  any  $h$-uniform
 subhypergraph   $K_{h}(m)$ with $V_1<\ldots <V_h$ is called an
  \emph{ordered copy of
$K_{h}(m)$} in $H$. Thus, the ordered version of the
theorem from Erd\"os and Simonovits about supersaturated uniform
hypergraphs \cite[Corollary 2, page 184]{ErdosS-82} can be stated
as follows.

\begin{theo}\label{ErdosS theorem}{\rm (Ordered Erd\"os-Simonovits)}
Given $c>0$ and two  integers $h \ge 2$ and $m \ge 1 $,  there
exist
 $c'>0$ and $N$  such that for all integers $n\ge N$,  if $H$ is a $h$-uniform
hypergraph  over $n$ vertices  having  at
least $c\cdot {n\choose h} $  hyperedges  then $H$ contains at least $c'n^{hm}$
ordered copies of $K_{h}(m)$.
\end{theo}

\noindent We will also use the following observation made when one consider an assignment of two colours, say $0$ and $1$,  to the hyperedges of $K_{h}(m)$. First let's say that  a vertex $s$  is  $c$-marked if $s$ belongs to at least one $c$-colored  hyperedge.  A subset   of vertices $S$  is  said $c$-marked  if any $s$  in $S$ is $c$-marked.   

\begin{clai}\label{marked}  Let $h \ge 2$,  $m \ge 1 $, and  $V_1, \cdots, V_{h}$   the   partition  associated  to $K_{h}(m)$. Consider an assignment of two colours to the     $m^{h}$ hyperedges of  $K_{h}(m)$, then at least one of the $V_i$ is marked.
 \end{clai}
 Indeed, suppose that  $V_1, \cdots, V_{h}$ are not $c$-marked. Now consider a vertex $s\in V_1$ then $s$ is $(1-c)$ marked else by construction of $K_{h}(m)$, $V_i$ would be $c$-marked  for all $i\geq 2$. Hence $V_1$ becomes $(1-c)$-marked.\\

Now let us show (\ref{fromCD3}), in other words  that  for any $K$-CNF formula $F$ such that   
${\cA}_{\Delta, \Delta'}(F)$ is dense then    ${\cB}_{\Delta, \Delta'}(F)$ is also dense.
For more readability  we will restrict our attention to the special case $K=3$, in using the above fact the proof  will be easily extendable to  any $K\geq 3$. Suppose there exist $\Theta(N^2)$ clauses in ${\cA}_{\Delta, \Delta'}(F)$ then, by the pigeon hole principle, at least for one of the eight types of clause we can find  $\Theta(N^2)$ clauses of this type in
${\cA}_{\Delta, \Delta'}(F)$.  Suppose, for example, that    

$$\#  \bigl \{ C=\overline {x_{i_1}} \ \vee x_{i_2}\vee \ \overline {x_{i_3}}, \ 1\leq i_1<i_2<i_3\leq N,  i_1\leq t,  \ F \wedge C \in  Y^{\Delta, \Delta'}\bigr \} =\Theta(N^2).$$ 

 \noindent From well chosen elements in  ${\cA}_{\Delta, \Delta'}(F)$ we now exhibit  an element in ${\cB}_{\Delta, \Delta'}(F)$. We consider the graph $H(F)$ associated to formula $F$: the set of vertices is $\{1, \ldots, N\}$ and for each $C=\overline {x_{i_1}} \ \vee x_{i_2}\vee \ \overline {x_{i_3}} \in  {\cA}_{\Delta, \Delta'}(F)$ we create  an edge $\{i_2,i_3\}$. Let 
$(\vec \sigma,\vec \tau)$ be a SAT-$x$-pair for $F$, then either $\sigma \notin S(C)$  or  $\tau \notin S(C)$.   Now,  following a fixed ordering  on the set of pairs of thruth assignments  we put the colour $0$ on the non colored edge  $\{i_2,i_3\}$ if $\sigma_{i_2}=0$ and $\sigma_{i_3}=1$ else we put the color $1$, having in this case $\tau_{i_2}=0$ and $\tau_{i_3}=1$.
Now,  let's take an ordered copy of $K(3,3)$ in $H(F)$ with partition $A=\{j_1, j_2, j_3 \}$ and $B=\{j_4, j_5, j_6 \}$. From Fact \ref{marked} we know that one part, say $A$,  is marked. In such a case  we have  $\sigma_{j_1}=0,\sigma_{j_2}=0, \sigma_{j_3}=0$ (A is $0$-marked) or  $\tau_{j_1}=0,\tau_{j_2}=0, \tau_{j_3}=0$ (A is $1$-marked)  hence  $(\vec \sigma,\vec \tau)$ is no longer  a SAT-$x$-pair for $F\wedge  (x_{j_1} \ \vee x_{j_2}\vee \  x_{j_3})$. If $B$ is marked then $(\vec \sigma,\vec \tau)$ is no longer  a SAT-$x$-pair for $F\wedge  ( \overline {x_{j_4}} \ \vee  \overline {x_{j_5}}\vee \   \overline {x_{j_6}})$.  Thus in any case $\{ ( x_{j_1} \ \vee x_{j_2}\vee \  x_{j_3} ), (\overline {x_{j_4}} \ \vee \overline {x_{j_5}}\vee \  \overline {x_{j_6}})\} \in {\cB}_{\Delta, \Delta'}(F)$.

\noindent By hypothesis $H(F)$ is a dense graph  so  from  Theorem \ref {ErdosS theorem} we can find $\Theta(N^6)$ copies of
$K(3,3)$ in $H(F)$. The above construction  provide $\Theta(N^6)$ elements in  $ {\cB}_{\Delta, \Delta'}(F)$ thus proving that this set is also dense.\end{proof}

\subsection{A general sharpness result}

Note that the above proof does not use any information about the shape of the set $\cD$ defining the $x$-unsatisfiability in terms of a subset of $\{0,\ldots, N\}$, namely the interval $ [Nx - \varepsilon(N), Nx + \varepsilon(N) ]$ (see $\eqref{subsetD}$). Actually we can consider  properties defined by  a non empty proper subset of $\{0,\ldots, N\}$ and we have proved  the following   general result:

\begin{theo}\label {generalsharp}
 Let $J_N$ be a non empty subset of $\{0, \ldots, N \}$ and consider  
 $${\cD_{J}} =\Bigl \{ (\vec{\sigma}, \vec{\tau}) \in \{0,1\} ^N \times \{0,1\} ^N \quad s. t.\quad  d_{\vec{\sigma} \vec{\tau}} \in J_N \Bigr \}.$$
Let $K\geq 3$ and   $Y_{J}$ be the set of $K$-CNF formula defined as:
 $$F \in Y_{J} \Longleftrightarrow  S(F) \times S(F) \quad \cap \quad {\cD_{J}} = \emptyset.$$
Then, $Y_J$ is a monotone property exhibiting a sharp threshlold.  

 \end{theo}

On one hand, any  upper bound for the satisfiability threshold, for instance $\eqref{cond0OK}$,  is  an upper bound for all $Y_J$ threshold. On the other hand, lemma \ref{simple} tells us  that a non complex  formula does not belongs to $Y_J$. Then, from \cite{KL-02}, we know that w.h.p a formula  whose ratio between the number of clauses and the number of varibles is less than $1/ K(K-1)$,  has   no complex component.   Thus it  provides a lower bound for all $Y_J$ threshold.

\section{Discussion and Conclusion}\label{conclu}
We have developed a simple and rigorous probabilistic method which is a first step towards a complete characterization of the clustered hard-SAT phase in the random satisfiability problem.
Our result is consistent with the clustering picture and supports the validity of the one-step replica symmetry breaking scheme of the cavity method for $K\geq 8$.

The study of $x$-satisfiability has the advantage that it
does not rely on a precise definition of clusters. 
Indeed, it is important to stress that the ``appropriate'' definition for clusters may vary according to the problem at hand. The natural choice seems to be the connected components of the space of SAT-assignments, where two adjacent assignments have by definition Hamming distance $1$. However, although this naive definition seems to work well on the satisfiability problem, it raises major difficulties on some other problems.
For instance, in $q$-colorability, it is useful to permit color exchanges between two adjacent vertices in addition to single-vertex color changes. In XORSAT, the naive definition is inadequate, since jumps from solution to solution can involve a large, yet finite, Hamming distance due to the hard nature of linear Boolean constraints \cite{MontanariSemerjian05-2}.

On the other hand, the existence of a gap in the $x$-satisfiability property is stronger than the original clustering hypothesis.
Clusters are expected to have a typical size, and to
be separated by a typical distance. However, even for typical formulas, there
exist atypical clusters, the sizes and separations of which may differ from
their typical values. Because of this variety of cluster sizes and
separations, a large range of distances is available to pairs of
SAT-assignments, which our $x$-satisfiability analysis takes into account.
What we have shown suggests that, for typical formulas, the maximum size of all
clusters is smaller than the minimum distance between two clusters (for a
certain range of $\alpha$ and $K\geq 8$). This is a sufficient condition for
clustering, but by no means a necessary one. As a matter of fact, our
large $K$ analysis conjectures that $\alpha_1(K)$ (the smaller $\alpha$ such that
Conjecture \ref{cluster} is verified) scales as $2^{K-1} \ln 2$, whereas
$\alpha_d(K)$ (where the replica symmetry breaking occurs) and $\alpha_s(K)$
(where the one-step RSB Ansatz is supposed to be valid) scale as $2^K\ln K/K$
\cite{MMZ-RSA}. 
According to the physics interpretation, in the range $\alpha_s(K)<a<\alpha_1(K)$, there exist clusters, but they are not detected by the $x$-satisfiability approach.
This limitation might account for the failure of our
method for small values of $K$ ---\,even though more sophisticated techniques
for evaluating the $x$-satisfiability threshold $\alpha_c(K,x)$ might yield some
results for $K<8$. Still, the conceptual simplicity of our method makes it a
useful tool for proving similar phenomena in other systems of computational or
physical interest.

A better understanding of the structure of the space of SAT-assignments could
be gained by computing the average configurational entropy of pairs of clusters at fixed distance, which contains details about how intra-cluster sizes and inter-cluster distances are distributed. This would yield the value of the $x$-satisfiability threshold.
Such a computation was carried out at a heuristic level within the framework of the cavity method for the random XORSAT problem \cite{MoraMezard06}, and should be extendable to the satisfiability problem or to other CSPs.

This work has been supported in part by the EC through the network 
MTR 2002-00319 `STIPCO' and the FP6 IST consortium `EVERGROW'.

This paper, signed in alphabetic order, is based on previous work by Mora M\'ezard and Zecchina reported in Sec.~1--4, 6. The proof in Sec.~5 is due to Daud\'e.

\bibliographystyle{unsrt}

\end{document}